\documentstyle[12pt,preprint,tighten,floats,aps,psfig]{revtex}
\def\simlt{\stackrel{<}{{}_\sim}}
\def\simgt{\stackrel{>}{{}_\sim}}
\tighten
\catcode`\@=11\relax
\newwrite\@unused
\def\typeout#1{{\let\protect\string\immediate\write\@unused{#1}}}
\def\figurepath{./}

\def\@nnil{\@nil}
\def\@empty{}
\def\@psdonoop#1\@@#2#3{}
\def\@psdo#1:=#2\do#3{\edef\@psdotmp{#2}\ifx\@psdotmp\@empty \else
    \expandafter\@psdoloop#2,\@nil,\@nil\@@#1{#3}\fi}
\def\@psdoloop#1,#2,#3\@@#4#5{\def#4{#1}\ifx #4\@nnil \else
       #5\def#4{#2}\ifx #4\@nnil \else#5\@ipsdoloop #3\@@#4{#5}\fi\fi}
\def\@ipsdoloop#1,#2\@@#3#4{\def#3{#1}\ifx #3\@nnil 
       \let\@nextwhile=\@psdonoop \else
      #4\relax\let\@nextwhile=\@ipsdoloop\fi\@nextwhile#2\@@#3{#4}}
\def\@tpsdo#1:=#2\do#3{\xdef\@psdotmp{#2}\ifx\@psdotmp\@empty \else
    \@tpsdoloop#2\@nil\@nil\@@#1{#3}\fi}
\def\@tpsdoloop#1#2\@@#3#4{\def#3{#1}\ifx #3\@nnil 
       \let\@nextwhile=\@psdonoop \else
      #4\relax\let\@nextwhile=\@tpsdoloop\fi\@nextwhile#2\@@#3{#4}}
\def\psdraft{
	\def\@psdraft{0}
}
\def\psfull{
	\def\@psdraft{100}
}
\psfull
\newif\if@prologfile
\newif\if@postlogfile
\newif\if@noisy
\def\pssilent{
	\@noisyfalse
}
\def\psnoisy{
	\@noisytrue
}
\psnoisy
\newif\if@bbllx
\newif\if@bblly
\newif\if@bburx
\newif\if@bbury
\newif\if@height
\newif\if@width
\newif\if@rheight
\newif\if@rwidth
\newif\if@clip
\newif\if@verbose
\def\@p@@sclip#1{\@cliptrue}

\def\@p@@sfile#1{\def\@p@sfile{null}%
	        \openin1=#1
		\ifeof1\closein1%
		       \openin1=\figurepath#1
			\ifeof1\typeout{Error, File #1 not found}
			\else\closein1
			    \edef\@p@sfile{\figurepath#1}%
                        \fi%
		 \else\closein1%
		       \def\@p@sfile{#1}%
		 \fi}
\def\@p@@sfigure#1{\def\@p@sfile{null}%
	        \openin1=#1
		\ifeof1\closein1%
		       \openin1=\figurepath#1
			\ifeof1\typeout{Error, File #1 not found}
			\else\closein1
			    \def\@p@sfile{\figurepath#1}%
                        \fi%
		 \else\closein1%
		       \def\@p@sfile{#1}%
		 \fi}

\def\@p@@sbbllx#1{
		\@bbllxtrue
		\dimen100=#1
		\edef\@p@sbbllx{\number\dimen100}
}
\def\@p@@sbblly#1{
		\@bbllytrue
		\dimen100=#1
		\edef\@p@sbblly{\number\dimen100}
}
\def\@p@@sbburx#1{
		\@bburxtrue
		\dimen100=#1
		\edef\@p@sbburx{\number\dimen100}
}
\def\@p@@sbbury#1{
		\@bburytrue
		\dimen100=#1
		\edef\@p@sbbury{\number\dimen100}
}
\def\@p@@sheight#1{
		\@heighttrue
		\dimen100=#1
   		\edef\@p@sheight{\number\dimen100}
}
\def\@p@@swidth#1{
		\@widthtrue
		\dimen100=#1
		\edef\@p@swidth{\number\dimen100}
}
\def\@p@@srheight#1{
		\@rheighttrue
		\dimen100=#1
		\edef\@p@srheight{\number\dimen100}
}
\def\@p@@srwidth#1{
		\@rwidthtrue
		\dimen100=#1
		\edef\@p@srwidth{\number\dimen100}
}
\def\@p@@ssilent#1{ 
		\@verbosefalse
}
\def\@p@@sprolog#1{\@prologfiletrue\def\@prologfileval{#1}}
\def\@p@@spostlog#1{\@postlogfiletrue\def\@postlogfileval{#1}}
\def\@cs@name#1{\csname #1\endcsname}
\def\@setparms#1=#2,{\@cs@name{@p@@s#1}{#2}}
\def\ps@init@parms{
		\@bbllxfalse \@bbllyfalse
		\@bburxfalse \@bburyfalse
		\@heightfalse \@widthfalse
		\@rheightfalse \@rwidthfalse
		\def\@p@sbbllx{}\def\@p@sbblly{}
		\def\@p@sbburx{}\def\@p@sbbury{}
		\def\@p@sheight{}\def\@p@swidth{}
		\def\@p@srheight{}\def\@p@srwidth{}
		\def\@p@sfile{}
		\def\@p@scost{10}
		\def\@sc{}
		\@prologfilefalse
		\@postlogfilefalse
		\@clipfalse
		\if@noisy
			\@verbosetrue
		\else
			\@verbosefalse
		\fi
}
\def\parse@ps@parms#1{
	 	\@psdo\@psfiga:=#1\do
		   {\expandafter\@setparms\@psfiga,}}
\newif\ifno@bb
\newif\ifnot@eof
\newread\ps@stream
\def\bb@missing{
	\if@verbose{
		\typeout{psfig: searching \@p@sfile \space  for bounding box}
	}\fi
	\openin\ps@stream=\@p@sfile
	\no@bbtrue
	\not@eoftrue
	\catcode`\%=12
	\loop
		\read\ps@stream to \line@in
		\global\toks200=\expandafter{\line@in}
		\ifeof\ps@stream \not@eoffalse \fi
		\@bbtest{\toks200}
		\if@bbmatch\not@eoffalse\expandafter\bb@cull\the\toks200\fi
	\ifnot@eof \repeat
	\catcode`\%=14
}	
\catcode`\%=12
\newif\if@bbmatch
\def\@bbtest#1{\expandafter\@a@\the#1
\long\def\@a@#1
\long\def\bb@cull#1 #2 #3 #4 #5 {
	\dimen100=#2 bp\edef\@p@sbbllx{\number\dimen100}
	\dimen100=#3 bp\edef\@p@sbblly{\number\dimen100}
	\dimen100=#4 bp\edef\@p@sbburx{\number\dimen100}
	\dimen100=#5 bp\edef\@p@sbbury{\number\dimen100}
	\no@bbfalse
}
\catcode`\%=14
\def\compute@bb{
		\no@bbfalse
		\if@bbllx \else \no@bbtrue \fi
		\if@bblly \else \no@bbtrue \fi
		\if@bburx \else \no@bbtrue \fi
		\if@bbury \else \no@bbtrue \fi
		\ifno@bb \bb@missing \fi
		\ifno@bb \typeout{FATAL ERROR: no bb supplied or found}
			\no-bb-error
		\fi
		\count203=\@p@sbburx
		\count204=\@p@sbbury
		\advance\count203 by -\@p@sbbllx
		\advance\count204 by -\@p@sbblly
		\edef\@bbw{\number\count203}
		\edef\@bbh{\number\count204}
}
\def\in@hundreds#1#2#3{\count240=#2 \count241=#3
		     \count100=\count240	
		     \divide\count100 by \count241
		     \count101=\count100
		     \multiply\count101 by \count241
		     \advance\count240 by -\count101
		     \multiply\count240 by 10
		     \count101=\count240	
		     \divide\count101 by \count241
		     \count102=\count101
		     \multiply\count102 by \count241
		     \advance\count240 by -\count102
		     \multiply\count240 by 10
		     \count102=\count240	
		     \divide\count102 by \count241
		     \count200=#1\count205=0
		     \count201=\count200
			\multiply\count201 by \count100
		 	\advance\count205 by \count201
		     \count201=\count200
			\divide\count201 by 10
			\multiply\count201 by \count101
			\advance\count205 by \count201
		     \count201=\count200
			\divide\count201 by 100
			\multiply\count201 by \count102
			\advance\count205 by \count201
		     \edef\@result{\number\count205}
}
\def\compute@wfromh{
		\in@hundreds{\@p@sheight}{\@bbw}{\@bbh}
		\edef\@p@swidth{\@result}
}
\def\compute@hfromw{
		\in@hundreds{\@p@swidth}{\@bbh}{\@bbw}
		\edef\@p@sheight{\@result}
}
\def\compute@handw{
		\if@height 
			\if@width
			\else
				\compute@wfromh
			\fi
		\else 
			\if@width
				\compute@hfromw
			\else
				\edef\@p@sheight{\@bbh}
				\edef\@p@swidth{\@bbw}
			\fi
		\fi
}
\def\compute@resv{
		\if@rheight \else \edef\@p@srheight{\@p@sheight} \fi
		\if@rwidth \else \edef\@p@srwidth{\@p@swidth} \fi
}
\def\compute@sizes{
	\compute@bb
	\compute@handw
	\compute@resv
}
\def\psfig#1{\vbox {
	\ps@init@parms
	\parse@ps@parms{#1}
	\compute@sizes
	\ifnum\@p@scost<\@psdraft{
		\if@verbose{
			\typeout{psfig: including \@p@sfile \space }
		}\fi
		\special{ps::[begin] 	\@p@swidth \space \@p@sheight \space
				\@p@sbbllx \space \@p@sbblly \space
				\@p@sbburx \space \@p@sbbury \space
				startTexFig \space }
		\if@clip{
			\if@verbose{
				\typeout{(clip)}
			}\fi
			\special{ps:: doclip \space }
		}\fi
		\if@prologfile
		    \special{ps: plotfile \@prologfileval \space } \fi
		\special{ps: plotfile \@p@sfile \space }
		\if@postlogfile
		    \special{ps: plotfile \@postlogfileval \space } \fi
		\special{ps::[end] endTexFig \space }
		\vbox to \@p@srheight true sp{
			\hbox to \@p@srwidth true sp{
				\hss
			}
		\vss
		}
	}\else{
		\vbox to \@p@srheight true sp{
		\vss
			\hbox to \@p@srwidth true sp{
				\hss
				\if@verbose{
					\@p@sfile
				}\fi
				\hss
			}
		\vss
		}
	}\fi
}}
\def\psglobal{\typeout{psfig: PSGLOBAL is OBSOLETE; use psprint -m instead}}
\catcode`\@=12\relax
\begin{document}
\draft
\title{Probing the phases in the MSSM via $e^{+}e^{-}\rightarrow h A$} 
\author{D. A. Demir\footnote{Present Address: ICTP, Trieste, Italy.}}
\address{Middle East Technical University, Department of Physics,
06531, Ankara, Turkey}
\date{\today}
\maketitle

\begin{abstract}
In the presence of the supersymmetric CP phases, 
$e^{+}e^{-}\rightarrow Z^{*} \rightarrow h A$ scattering is analyzed, with 
special emphasis on $Z^{*} h A$ vertex, taking into account radiatve 
corrections due to dominant top quark and top squark loops in the Higgs
sector as well as the vertex formfactors. It is found that the lightest 
Higgs remains essentially CP--even whereas the heavier ones mix with each other 
strongly. Moreover, the supersymmetric CP phases are found to create tree-level 
couplings between a pseudoscalar and identical sfermion mass eigenstates which,
together with the radiative corrections in the Higgs sector, 
are found to enhance the vertex formfactors significantly. 
\end{abstract}
\newpage
\section{Introduction}
The Minimal Supersymmetric Standard Model (MSSM) is an appealing extension
of the minimal Standard Model in that it resolves the well-known gauge 
hierarchy problem of the latter. The MSSM Higgs sector comprises two 
opposite hypercharge Higgs doublets whose all quartic couplings are 
fixed by the gauge couplings at the tree level. In the MSSM Lagrangian, 
other than Yukawa couplings and the $\mu$ parameter coming from the superpotential, 
there are several mass parameters in the soft supersymmetry breaking sector \cite{rosiek}.
These parameters consist of the gaugino masses $M_{a}$, sfermion mass-squareds 
$m_{\tilde{f}}^{2}$, Higgs-sfermion trilinear couplings $A_{f}$, 
Higgs mass-squareds $m_{i}^{2}$, and bilinear Higgs mixing mass $B\mu$, all of which could, 
in principle, be complex. However, not all these phases are physical \cite{phase1}, 
that is, an arbitrary physical quantity depends only on certain combinations of them as 
dictated by the global $U(1)$ symmetries of the MSSM Lagrangian \cite{phase2}.
Indeed, one can choose observable phases to be those of CKM matrix, trilinear couplings
$A_{f}$, and the $\mu$ parameter \cite{phase1}, and we will adopt this convention below.

For getting information about the physical implications of these phases the most convenient way is 
to look for the collider signatures of certain collision processes. Among others, the 
lepton colliders can provide a clean signature independent of the hadronic uncertainities, 
depending on the final state particles. In future generation of the colliders, with increasing
center of mass energy of the leptons it may be possible to detect supersymmetric particles, 
in particular, the Higgs particles \cite{NLC,muon}. In this work we discuss $e^{+}e^{-}\rightarrow h A$
scattering in the presence of explicit CP--violation in the MSSM Lagrangian through the complex 
$\mu$ and $A$ paarmeters. This process has been analyzed in detail in \cite{pokorski} at one-loop 
level, and radiative corrections were found to contribute $\sim 5\%$ in regions of the parameter space 
where the cross section is maximized. Other than this one-loop analysis, as emhasized by Haber \cite{decoupling},
$h A Z$ coupling becomes vanishingly small in large $\tan\beta-M_{A}$ regime. This process, compared 
to Bjorken process and $W$ or $Z$ fusion, involves two Higgs particles at the final state whose 
mixings and indefinite CP characteristics can affect the cross section whereby providing some 
information on the supersymmetry search. As will be seen below, mainly there are two distinct 
effects of these phases on the pair production process: (1) Mixing between the Higgs scalars 
due to CP-- violation in the Higgs sector, and (2) loop effects of the sfermions on the 
$h A Z$ vertex. Below, analytical results will be general; however, in the numerical analysis we will
assume a vanishing phase for $\mu$ in accordance with the EDM \cite{phase1,olive1} and cosmological 
constraints \cite{olive2} though several ways of relaxing these constraints have been suggested 
\cite{sidestep}. Recently, the supersymmetric CP phases have gotten much interest in both Higgs 
phenomenology \cite{ben,pilaftsis,wagner} and FCNC processes \cite{biz}.

\section{Supersymmetric CP Phases and $e^{+}e^{-}\rightarrow h A$ Scattering}
The supersymmetric CP phases ${\cal{A}}rg\{A_{f}\}$ and ${\cal{A}}rg\{\mu\}$ show 
up in sfermion, chargino and neutralino mass matrices so that the processes involving these
particles as well as their loop effects depend on these phases explicitly 
\cite{biz,ben,pilaftsis,wagner}. In the presence of these phases there is 
explicit CP--violation in the Higgs sector; namely, the mass eigenstate 
Higgs bosons are mixtures of scalars having different CP--properties.
Moreover, these phases enhance the radiative corrections to certain quantities by 
introducing novel interaction vertices not found in the CP--invariant theory. 
In what follows we shall discuss these points in detail when analyzing the effective 
scalar--pseudoscalar--$Z$ boson vertex. 

The loop effects of the MSSM particle spectrum on the Higgs potential could be parametrized 
in a simple and elegant way by applying the effective potential method \cite{effpot}. Its implicit
renormalization prescription corresponds to the $\overline{DR}$ scheme. Indeed, if the theory
is regularized by dimensional reduction (reducing to the usual dimensional regularization in the absence of 
gauge boson loops, and with the conventional algebra for Dirac matrices), in the Landau gauge, 
the effective potential is given by 
\begin{eqnarray}
V=V_{0}+\frac{1}{64 \pi^{2}} \mbox{Str} {\cal{M}}^{4} (\log \frac{{\cal{M}}^{2}}{Q^{2}} -\frac{3}{2}-\Delta)  
\end{eqnarray}
where $V_{0}$ is the tree level MSSM potential depending on the bare parameters, 
$\Delta=2/(4-D)-\gamma+\log 4 \pi$, $D$ is the dimension of spacetime, $\gamma$ is Euler constant,
${\cal{M}}^{2}$ is the field dependent mass-squared matrix of all fields, and $Q$ is the renormalization scale. 
Here the renoralization scale is naturally at the weak scale and we shall take $Q^{2}\sim m_{t}^{2}$ though 
slightly different choices are also possible \cite{effpot}. Expressing the bare parameters and fields in 
$V_{0}$ in terms of the renormalized ones and one-loop counterterms as 
$V_{0}=V_{tree}+\Delta V$, and using the $minimal$ prescription $\Delta V=\Delta \mbox{Str} {\cal{M}}^{4}/(64 \pi^{2})$, 
the UV divergence of (1) is cured. Here $V_{tree}$ is the tree-level MSSM Higgs potential composed of 
renormalized parameters and fields only. As usual, the second order derivatives of the effective potential 
(1) with respect to the components of the Higgs fields, evaluated at the stationarity point, give the Higgs 
mass-squared matrix from which the radiatively--corrected Higgs masses and Higgs mixing matrix follow. 
The effective potential (1) gets the most important contributions from the top quark and top squark 
loops for moderate values of $\tan\beta$ (for $\tan\beta\sim 60$ bottom quark-squark, tau lepton-slepton contributions
become important) \cite{effpot}. Therefore, to a good approximation, it is sufficient to take into 
account only the dominant top quark and top squark loops, and neglect the contributions of other particles as justified by the
analysis of \cite{justi}. Neglecting the D-term contributions, in $(\tilde{t}_{L},
\tilde{t}_{R})$ basis field-- dependent stop mass-squared matrix is given by 
\begin{eqnarray}
M_{\tilde{t}}=\left(\begin{array}{c c} M_{\tilde{L}}^{2}+m_{t}^{2} &
h_{t}(A_{t}H^{0}_{2}-\mu^{*}{H_{1}^{0}}^{*})\\
h_{t}(A_{t}^{*}{H^{0}_{2}}^{*}-\mu H_{1}^{0}) & M_{\tilde{R}}^{2}+m_{t}^{2}
\end{array}\right)
\end{eqnarray}
where the top quark mass has the usual expression $m_{t}^{2}=h_{t}^{2}{|H_{2}^{0}|}^{2}$, and 
$M_{\tilde{L},\tilde{R}}^{2}$ are the soft mass parameters of left-- and right-- handed stop fields, 
respectively. Here $H_{1,2}^{0}$ are the neutral components of the Higgs doublets, 
and in the true vacuum state of the MSSM, $<H_{2}^{0}>=v_{2}/\sqrt{2}$, 
$<H_{1}^{0}>=v_{1}/\sqrt{2}$ such that $v_{1}^{2}+v_{2}^{2}=4 M_{W}^{2}/g^{2}$, and 
$v_{2}/v_{1}\equiv \tan\beta$. In this vacuum state stop mass-squared matrix
(1) is diagonalized as $U_{\tilde{t}}^{\dagger} M_{\tilde{t}} U_{\tilde{t}}= \mbox{diag}
(m_{\tilde{t}_{1}}^{2}, m_{\tilde{t}_{2}}^{2})$ via the unitary matrix
\begin{eqnarray}
U_{\tilde{t}}=\left (\begin{array}{c c}
\cos\theta_{\tilde{t}}&
\sin\theta_{\tilde{t}}e^{i\gamma_{t}}\\
-\sin\theta_{\tilde{t}}e^{-i\gamma_{t}}&
\cos\theta_{\tilde{t}}\end{array}\right)
\end{eqnarray}
whose entries have the meaning 
\begin{eqnarray}
\tan 2\theta_{\tilde{t}}&=& -\frac{2 m_{t}\tilde{A}_{t}}{m_{L}^{2}-m_{R}^{2}}\nonumber\\
\tan\gamma_{t}&=&\frac{|A_{t}|\sin\gamma_{A}+|\mu|\cot\beta \sin\gamma_{\mu}}
{|A_{t}|\cos \gamma_{A}-|\mu|\cot\beta \cos\gamma_{\mu}} \nonumber\\
\tilde{A}_{t}&=&[|A_{t}|^{2}+|\mu|^{2}\cot^{2}\beta-2|A_{t}||\mu|\cot\beta
\cos(\gamma_{\mu}+\gamma_{A})]^{1/2}
\end{eqnarray}
where $\gamma_{A}\equiv {\cal{A}}rg\{A_{f}\}$, $\gamma_{\mu}\equiv {\cal{A}}rg\{\mu\}$. The mass eigenstate stops
are denoted by $\tilde{t}_{1}$ and $\tilde{t}_{2}$, with $m_{\tilde{t}_{1}}< m_{\tilde{t}_{2}}$,  where
\begin{eqnarray}
m_{\tilde{t}_{1}(\tilde{t}_{2})}=(1/2)(2 m_{t}^{2}+ M_{\tilde{L}}^{2}+
M_{\tilde{R}}^{2}-(+)\sqrt{(M_{\tilde{L}}^{2}-M_{\tilde{R}}^{2})^{2}+4 m_{t}^{2}\tilde{A}_{t}^{2}})\;.
\end{eqnarray}
Since $\tilde{A}_{t}$ increases as $\gamma_{A}+\gamma_{\mu}$ increases from zero to $\pi$, $m_{\tilde{t}_{1}}$
($m_{\tilde{t}_{2}}$) decreases (increases) with $\gamma_{A}+\gamma_{\mu}$. Therefore, $m_{\tilde{t}_{1}}$ becomes 
minimal for CP phases around $\pi$. This point will be useful in discussing the radiative corrections to 
scalar--pseudoscalar--$Z$ boson vertex. 

For computing the radiative corrections to Higgs masses and mixings one applies the usual procedure, that is, 
the stop mass--squared matrix (2) is diagonalized to obtain field dependent stop masses which are inserted to the
effective potential formulae (1) together with the field dependent mass for the top quark. Then the second order
derivatives of (1) evaluated at the stationarity point give the radiatively--corrected Higgs mass-squared matrix.
One can find detailed expressions for the elements of Higgs mass-squared matrix in \cite{ben} 
which uses the basis ${\cal{B}}=({\cal{R}}e[H_{1}^{0}], {\cal{R}}e[H_{2}^{0}], A\equiv \sin\beta {\cal{I}}m[H_{1}^{0}] + 
\cos\beta {\cal{I}}m[H_{2}^{0}])$. Diagonalization of the Higgs mass-squared matrix produces three mass-eigenstate scalars
$H_{1}$, $H_{2}$, $H_{3}$ which are choosen to correspond to $h$, $H$ and $A$ at $\gamma_{A}= 0$, 
respectively. Symbolically one has
\begin{eqnarray}
\left(\begin{array}{c} H_{1}\\ H_{2}\\ H_{3} \end{array}\right)= \left(\begin{array}{c c c} R_{11} & R_{12} & R_{13}\\
 R_{21} & R_{22} & R_{23}\\  R_{31} & R_{32} & R_{33}\end{array}\right)\left(\begin{array}{c} {\cal{R}}e[H_{1}^{0}]\\ 
{\cal{R}}e[H_{2}^{0}]\\A \end{array}\right)\;.
\end{eqnarray}
From this matrix equality it is obvious that the mass-eigenstate scalars $H_{i}$ are no longer CP eigenstates. 
This CP--breaking follows from the associated entries of the Higgs mass-squared matrix which are all 
proportional to $\sin (\gamma_{A}+\gamma_{\mu})$; namely, CP--violation in the Higgs sector is lifted once
the supersymmetric CP--phases vanish \cite{ben}. Elements of the matrix $R$ characterize the CP--violation in the Higgs 
sector and they appear in couplings of $H_{i}$ to fermions, gauge bosons, and other Higgs particles as well
\cite{ben,wagner}. To compute the radiative corrections to scalar--pseudoscalar--$Z$ boson vertex in the presence of
the supersymmetric CP--phases one needs Feynman rules for certain vertices. The $Z$ boson couples to Higgs particles 
and stops as follows:
\begin{eqnarray}
V_{Z H_{i} H_{j}}&=&(\cos\beta R_{j2}-\sin\beta R_{j1})R_{i3}-(\cos\beta R_{i2}-\sin\beta R_{i1})R_{j3}\nonumber\\
V_{Z\tilde{t}_{1}\tilde{t}_{1}}&=&-i(\frac{1}{2}\cos^{2}\theta_{\tilde{t}}-\frac{2}{3}s_{W}^{2})\nonumber\\
V_{Z\tilde{t}_{1}\tilde{t}_{2}}&=&-i\frac{1}{4}\sin 2\theta_{\tilde{t}} e^{i\gamma_{t}}\\
V_{Z\tilde{t}_{2}\tilde{t}_{1}}&=&-i\frac{1}{4}\sin 2\theta_{\tilde{t}} e^{-i\gamma_{t}}\nonumber\\
V_{Z\tilde{t}_{2}\tilde{t}_{2}}&=&-i(\frac{1}{2}\sin^{2}\theta_{\tilde{t}}-\frac{2}{3}s_{W}^{2})\nonumber
\end{eqnarray}
where $i,j=1,2,3$. For future convenience, $V_{Z\tilde{t}_{1}\tilde{t}_{1}}$ is given in units of
$G\equiv\sqrt{g^{2}+{g'}^{2}}$, and $V_{Z H_{i} H_{j}}$ in units of $G/2$ with their Lorentz 
structures suppressed. On the other hand, the CP--odd componenet of $H_{i}$ couples to stops as
follows
\begin{eqnarray}
V_{H_{i}\tilde{t}_{1}\tilde{t}_{1}}^{(P)}&=&\frac{1}{2}\sin 2 \theta_{\tilde{t}}( A_{P}^{i}-{A_{P}^{i}}^{*})\nonumber\\
V_{H_{i}\tilde{t}_{2}\tilde{t}_{2}}^{(P)}&=&-\frac{1}{2}\sin 2 \theta_{\tilde{t}}(A_{P}^{i}-{A_{P}^{i}}^{*})\\
V_{H_{i}\tilde{t}_{1}\tilde{t}_{2}}^{(P)}&=&\sin^{2}\theta_{\tilde{t}} A_{P}^{i} e^{i\gamma_{t}}+\cos^{2}\theta_{\tilde{t}}
{A_{P}^{i}}^{*}e^{-i\gamma_{t}}\nonumber\\
V_{H_{i}\tilde{t}_{2}\tilde{t}_{1}}^{(P)}&=&-(\cos^{2}\theta_{\tilde{t}} A_{P}^{i} e^{-i\gamma_{t}}+\sin^{2}\theta_{\tilde{t}}
{A_{P}^{i}}^{*}e^{-i\gamma_{t}})\nonumber
\end{eqnarray} 
where $A_{P}^{i}=(|A_{t}|\cos\beta e^{i\gamma_{A}}+|\mu|\sin\beta e^{-i\gamma_{\mu}})e^{i\gamma_{t}} R_{i 3}$ shows the 
effects of the stop left-right mixings on the couplings of the CP--odd component of $H_{i}$.
Finally, the CP--even component of $H_{i}$ couples to stops via 
\begin{eqnarray}
V_{H_{i}\tilde{t}_{1}\tilde{t}_{1}}^{(S)}&=&i\{\frac{4}{3}\frac{M_{Z}^{2}}{m_{t}}s_{W}^{2}(1+\frac{3-8 s_{W}^{2}}{4
s_{W}^{2}}\cos^{2}\theta_{\tilde{t}})\sin\beta (\cos\beta R_{i1}-\sin\beta R_{i2})+2 m_{t} R_{i 2}\nonumber\\
&-&\frac{1}{2}\sin 2 \theta_{\tilde{t}}( A_{S}^{i}+{A_{S}^{i}}^{*})\}\nonumber\\
V_{H_{i}\tilde{t}_{2}\tilde{t}_{2}}^{(S)}&=&i\{\frac{4}{3}\frac{M_{Z}^{2}}{m_{t}}s_{W}^{2}(1+\frac{3-8 s_{W}^{2}}{4
s_{W}^{2}}\sin^{2}\theta_{\tilde{t}})\sin\beta (\cos\beta R_{i1}-\sin\beta R_{i2})+2 m_{t} R_{i 2}\nonumber\\
&+&\frac{1}{2}\sin 2 \theta_{\tilde{t}}( A_{S}^{i}+{A_{S}^{i}}^{*})\}\\
V_{H_{i}\tilde{t}_{1}\tilde{t}_{2}}^{(S)}&=&i\{\frac{1}{6}\frac{M_{Z}^{2}}{m_{t}}(3-8 s_{W}^{2})
\sin 2\theta_{\tilde{t}}\sin\beta(\cos\beta R_{i1}-\sin\beta R_{i2})e^{\gamma_{t}}\nonumber\\
&+&\cos^{2}\theta_{\tilde{t}}{A_{S}^{i}}^{*}e^{i\gamma_{t}}-\sin^{2}\theta_{\tilde{t}}
A_{S}^{i} e^{i\gamma_{t}}\}\nonumber\\
V_{H_{i}\tilde{t}_{2}\tilde{t}_{1}}^{(S)}&=&i\{\frac{1}{6}\frac{M_{Z}^{2}}{m_{t}}(3-8 s_{W}^{2})
\sin 2\theta_{\tilde{t}}\sin\beta(\cos\beta R_{i1}-\sin\beta R_{i2})e^{-\gamma_{t}}\nonumber\\
&+&\cos^{2}\theta_{\tilde{t}}{A_{S}^{i}} e^{-i\gamma_{t}}-\sin^{2}\theta_{\tilde{t}}
{A_{S}^{i}}^{*}e^{-i\gamma_{t}}\}\nonumber
\end{eqnarray}
where $A_{S}^{i}=(|A_{t}|R_{i 2} e^{i \gamma_{A}}-|\mu|R_{i 1} e^{-i \gamma_{\mu}})e^{i \gamma_{t}}$ summarizes
nothing but the effects of stop left-right mixings on the CP--even component of $H_{i}$. All $H_{i}$ to
$\tilde{t}_{a}\tilde{t}_{b}$ couplings listed in (8) and (9) are given in units of $h_{t}/\sqrt{2}$ for future
convenience. 

Couplings of the Higgs scalars to stops depend on several parameters coming from the mass-squared matrices of stops
and Higgs scalars. Among all, stop left-right mixing angle $\theta_{\tilde{t}}$, CP--breaking phases $\gamma_{A,t}$
and Higgs mixing parameters $R_{ij}$ are particularly interesting. The CP--violating 
supersymmetric phases enter all couplings in (7)-(9), masses of the Higgs scalars and stops, and Higgs  mixing matrix $R$. 
These phases not only modify the couplings existing in the CP--invariant theory, but also create new ones as the
expressions of $V_{H_{i}\tilde{t}_{1}\tilde{t}_{1}}^{(P)}$ and $V_{H_{i}\tilde{t}_{2}\tilde{t}_{2}}^{(P)}$ show
explicitly. These two couplings are created solely by the supersymmetric CP phases and necessarily vanish in the
CP--conserving limit: $\gamma_{A,\mu}\rightarrow 0$. 

When the off-diagonal elements of the stop mass-squared matrix are large (provided the light stop mass in above 
the present LEP lower bound of $\sim 75\; \mbox{GeV}$) compared to its diagonal elements, the stop mixing becomes 
maximal, that is,  $\sin 2\theta_{\tilde{t}}\leadsto 1$. In this limit, $Z$ to $\tilde{t}_{a}\tilde{t}_{b\neq a}$,
$H_{i}^{P}$ to $\tilde{t}_{a}\tilde{t}_{b}$, and $H_{i}^{S}$ to $\tilde{t}_{a}\tilde{t}_{a}$ couplings are maximized 
as seen from (7)-(9). Of course such statements depend on relative strengths of $|A_{t}|$ and $|\mu|$ as well as 
their phases. In essence, one needs large chiral mixings to highlight the effects of novel $H_{i}^{P}\tilde{t}_{a}\tilde{t}_{a}$
couplings; however, in this limit $H_{i}^{S}\tilde{t}_{a}\tilde{t}_{a}$ and $H_{i}^{P}\tilde{t}_{a}\tilde{t}_{b\neq a}$
couplings (which exist in the CP--conserving limit too ) become also large though the corresponding formfactors are
expected to be suppressed partially by heavy $\tilde{t}_{2}$. The CP--compositions of the Higgs scalars are dictated by $R_{i j}$;
hence, whatever the strengths of the one-loop vertex corrections, they constitute the envelope of the effective
scalar--pseudoscalar--$Z$ boson vertices. These points will be clearer when the explicit numerical computation is 
carried out. 
  
For a proper analysis of the Higgs pair production, it is not sufficent to compute the radiative corrections to Higgs masses and
couplings, one has to compute also the radiative corrections to  scalar--pseudoscalar--$Z$ boson vertex. Once the radiative
corrections are switched on vector boson self energies as well as the vertex formfactors need be computed with the inclusion of
the entire MSSM particle spectrum \cite{pokorski}. However, consistent with the description of the Higgs sector, dominant
corrections come from the top quark and top squark loops. Using the form of the radiatively--corrected cross 
section \cite{pokorski} one can easily incorporate the radiative corrections to `tree' vertex 
$V_{Z H_{i} H_{j}}$ as follows
\begin{eqnarray}
\hat{V}_{Z H_{i} H_{j}}&=&V_{Z H_{i} H_{j}}+\beta_{h_{t}}\{\cos\beta
(R_{i2}R_{j3}-R_{j2}R_{i3}){\cal{Q}}_{t}(-p_{i},p_{j},m_{t},m_{t},m_{t})\nonumber\\&+&
(V_{Z\tilde{t}_{b}\tilde{t}_{c}}V_{H_{i}\tilde{t}_{a}\tilde{t}_{b}}^{(S)}V_{H_{j}\tilde{t}_{c}\tilde{t}_{a}}^{(P)}-
V_{Z\tilde{t}_{c}\tilde{t}_{b}}V_{H_{i}\tilde{t}_{b}\tilde{t}_{a}}^{(P)}V_{H_{j}\tilde{t}_{a}\tilde{t}_{c}}^{(S)})
{\cal{Q}}_{\tilde{t}}(-p_{i},p_{j},m_{\tilde{t}_{a}},m_{\tilde{t}_{b}},m_{\tilde{t}_{c}})\}
\end{eqnarray}
where $\beta_{h_{t}}=3 h_{t}^{2}/(16 \pi^{2})$, summation of $a,b=1,2$ is implied, and top quark and top squark triangles are
represented by the loop functions ${\cal{Q}}_{t}(-p_{i},p_{j},m_{t},m_{t},m_{t})$ and 
${\cal{Q}}_{\tilde{t}}(-p_{i},p_{j},m_{\tilde{t}_{a}},m_{\tilde{t}_{b}},m_{\tilde{t}_{c}})$, 
respectively. These vertex formfactors could be expressed in terms of the standard Veltman-Passarino 
loop functions \cite{veltman} as follows
\begin{eqnarray}
{\cal{Q}}_{t}(p, q, m, m, m)&=& B_{0}(p-q, m)+2 m^{2} C_{0}(p,q, m, m, m)+p^{2} C_{1}(p, q, m, m, m)\nonumber\\&+&
q^{2} C_{2}(p, q, m, m, m)\nonumber\\
{\cal{Q}}_{\tilde{t}}(p, q, m_{a}, m_{b}, m_{c})&=&-2(C_{0}(p,q, m_{a}, m_{b}, m_{c}) + C_{1}(p, q, m_{a}, m_{b}, m_{c})
\nonumber\\&+& C_{2}(p, q, m_{a}, m_{b}, m_{c}))
\end{eqnarray}
where the notation and definitions of \cite{denner} are adopted. Here $p_{i}^{2}=M_{H_{i}}^{2}$,
$p_{j}^{2}=M_{H_{j}}^{2}$ and  $2 p_{i}.p_{j}= s-M_{H_{i}}^{2}-M_{H_{j}}^{2}$. The light stop contribution is finite whereas the 
top quark contribution, due to $B_{0}$ function, has a UV divergence, $\Delta$, which is renormalized with minimal 
prescription as was done for the effective potential (1). Moreover, $B_{0}$ has a scale dependence through 
$\log{m_{t}^{2}/Q^{2}}$.

There are several aspects of (10) deserving a detailed discussion. First, one notes that top quark contribution is proportional
to $\cos\beta$, which means that this contribution is suppressed for large $\tan\beta$ (before bottom and tau Yukawa coupligs
become comparable to top Yukawa coupling). In the same way $V_{Z H_{i} H_{j}}$ gets diminished for large $\tan\beta$
\cite{decoupling,hunter} especially when $|R_{i3}|$ is negligably small (this will be seen to hold in $H_{1}H_{j\neq 1}$
production). Unlike $V_{Z H_{i} H_{j}}$ and top quark loop contribution, however, the stop contribution is not necessarily
suppressed in the large $\tan\beta$ regime due to the fact that it has both $\sin\beta$ and $\cos\beta$ dependencies weigted by 
$|A_{t}|$ and $|\mu|$ as seen from the expression of $A_{P}^{i}$ below (8). In this way one expects stop contributions to lead 
important variations in the effective vertex depending on the supersymmetric parameter space adopted. From eq. (8) one observes
that the coupling of $H_{i}^{P}$ to $\tilde{t}_{1}\tilde{t}_{1}$ pair is purely imaginary so that its contribution to 
$|\hat{V}_{Z H_{i} H_{j}}|$ remains at the two-loop order unless 
${\cal{Q}}_{\tilde{t}}(-p_{i},p_{j},m_{\tilde{t}_{1}},m_{\tilde{t}_{1}},m_{\tilde{t}_{1}})$ develops an absorbtive part 
part entailing an interference with $V_{Z H_{i} H_{j}}$ and the dispersive part of the top quark contribution. These observations 
require light stop be light enough (weighing  $\sim m_{t}$) to have stop contributions be enmhanced.

To have better understanding of the role of the supersymmetric phases on Higgs pair production it is convenient to analyze
the one-loop vertex (10) numerically. However, for this purpose it is necessary to have a detailed knowledge of the mixing matrix
$R$ to identify the CP--impurities of the Higgs bosons. In the numerical analysis below we shall adopt the following parameter 
values 
\begin{eqnarray}
M_{\tilde{L}}=M_{\tilde{R}}=500\; \mbox{GeV};\, |A_{t}|= 1.3\; \mbox{TeV};\, |\mu|=\mu=250\; \mbox{GeV};\, \tilde{M}_{A}=200\;
\mbox{GeV}
\end{eqnarray}
having in mind an $e^{+}e^{-}$ collider with $\sqrt{s}=500\; \mbox{GeV}$ (for example the recently planned TESLA facility
\cite{tesla}). Here $\tilde{M}_{A}$ is analogous to pseudoscalar mass $M_{A}$ in the CP--invariant theory \cite{effpot}, and its
definition can be found in \cite{ben}. In the analysis, phase of $A_{t}$, $\gamma_{A}$, is treated as a free variable whereas
$\gamma_{\mu}$ is set to zero. In each case $\tan\beta=2$ and $30$ are considered seperately to illustrate low and high
$\tan\beta$ cases, respectively. Depicted in Figures 1-3  are the $\gamma_{A}$ dependence of the percentage
compositions of $H_{1}$, $H_{2}$ and $H_{3}$. Fig. 1 shows $R_{11}^{2}$ (solid curve), $R_{12}^{2}$ (dashed curve), and
$R_{13}^{2}$ (dotted curve) for $\tan\beta=2$ (left panel) and $\tan\beta=30$ (right panel). As the figure suggests,
$\tan\beta=2$, CP--even components of $H_{1}$ oscillate between $\sim 45\%$ and $\sim 55\%$ in the entire $\gamma_{A}$ range
while its CP--odd component remains below $1\%$ everywhere. The right panel shows the $\gamma_{A}$ dependence of the same
quantities for $\tan\beta=30$, from which  it is seen that $R_{12}^{2}$ remains above $\sim 90\%$, and correspondingly  
$R_{11}^{2}$ is always below $\sim 10\%$ line. This rearrangement of the CP--even components results from the large value of
$\tan\beta$ as in the CP--invariant theory \cite{decoupling}. Similar to Fig. 1 the CP--odd composition of $H_{1}$ is rather
small, never exceeding $\sim 0.5\%$ level. That the lightest Higgs ($H_{1}$) remains essentially CP--even follows from the
decoupling between the heavy and light sectors in the MSSM Higgs sector \cite{pilaftsis,ben,wagner}. Indeed, if $\tilde{M}_{A}$
were lighter (say below $\sim 150\; \mbox{GeV}$ with a low enough $\tan\beta$) $H_{1}$ would have a larger CP=-1 composition. 

Fig. 2 shows $\gamma_{A}$ dependence of $R_{21}^{2}$ (solid curve), $R_{22}^{2}$ (dashed curve), and $R_{23}^{2}$
(dotted curve) for $\tan\beta=2$ (left panel) and $\tan\beta=30$ (right panel) . As is seen form Fig. 3, for
$\gamma_{A}\rightarrow 0$, $H_{2}$ becomes the usual heavy CP--even Higgs boson of the MSSM, its CP--even components start at
$\gamma_{A}=0$ in agreemnet with Fig. 1 (left panel), and completely vanish at $\gamma_{A}=\pi$ at which it becomes a completely
CP--odd Higgs boson. Indeed, $H_{2}$ assumes a definite CP property for $\sin\gamma_{A}\rightarrow 0$; however, it has opposite
CP quantum numbers for $\gamma_{A}=0$ and $\gamma_{A}=\pi$. Swithching to Fig. 2 right panel one observes an even stronger
CP--impurity for $H_{2}$: Except for $\gamma_{A}\simlt 1$ ( and of course $\gamma_{A}\simgt 5$) $H_{2}$ is almost a pure CP--odd
Higgs boson where its CP--even composition remains below $\sim 12\%$. As in Fig. 3 for $\gamma=\pi$ $H_{2}$ is purely CP--odd.
The strong CP--impurity of $H_{2}$ automatically implies a symmetric situation for $H_{3}$ (the would-be CP--odd boson of the
CP--invariant theory) as suggested by Fig. 3. Indeed $H_{2}$ and $H_{3}$ are complementary to each other in the sense that
$H_{3}$ is mostly CP--even in regions where $H_{2}$ is nearly CP--odd. The lesson drawn from Figs. 1-3 is that the lightest
Higgs $H_{1}$ keeps becoming, to a good approximation, CP--even while the other two Higgs scalars mix with each other
significantly depending on $\gamma_{A}$ and $\tan\beta$. In forming these graphs use has been made of the parameter set in (8) in
which $\tilde{M}_{A}$ is fixed to $200\;\mbox{GeV}$. Similar to $M_{A}$ of the CP--invariant theory $\tilde{M}_{A}$ determines
the masses of heavy scalars $H_{2}$ and $H_{3}$. Together with large $\tan\beta$ values, large $\tilde{M}_{A}$ values imply the
decoupling limit described in \cite{decoupling}. However, for collider applications (with $\sqrt{s}=500\;\mbox{GeV}$ as
considered here) it is necessary to keep $\tilde{M}_{A}$ around $200\;\mbox{GeV}$ to allow for pair production of the Higgs
scalars. In accordance with the results of \cite{ben}, on the average, $M_{H_{1}}=130 \;\mbox{GeV}\; (155 \;\mbox{GeV})$,
$M_{H_{2}}=210 \;\mbox{GeV}\; (200 \;\mbox{GeV})$, and $M_{H_{3}}=200 \;\mbox{GeV}\; (200 \;\mbox{GeV})$ for $\tan\beta=2 (30)$
in the entire range of $\gamma_{A}$. Therefore, with $\sqrt{s}=500\;\mbox{GeV}$, it is possible to produce even $H_{2}-H_{3}$
pairs despite strong kinematic suppression compared to $H_{1}-H_{2}$ or $H_{1}-H_{3}$ productions. 
   
The CP compositions of $H_{2}$ and $H_{3}$ shown in Figs. 2 and 3 need further discussion. One notices that $\gamma_{A}=0$ and
$\gamma_{A}=\pi$ are equivalent in the sense that CP--breaking components of the Higgs mass-squared matrix vanish at both points.
However, according to the CP compositions of $H_{2}$ and $H_{3}$ these points are no longer equivalent. This follows from the 
fact that the radiative corrections at these two points are no longer equivalent since the quantities involving $\cos \gamma_{A,\mu}$
(which remain non-vanishing in the CP--conserving limit) reverse their sign as one switches from $\gamma_{A}=0$ to $\gamma_{A}=\pi$.
Therefore, as an example, one observes from (5) that for $\gamma_{A}=\pi$ light stop (heavy stop) assumes its smallest (largest)
possible mass for a given parameter space. This then maximizes $\log m_{\tilde{t}_{2}}^{2}/m_{\tilde{t}_{1}}^{2}$ type stop--splitting 
contributions modifying the strength of the radiative corrections. Other than all these, diagonalization of the Higgs mass-squared
matrix (described in (6)) uses only the properties of the scalars at $\gamma_{A}=0$ in naming them, and does not care their behaviour
at finite $\gamma_{A}$.

Given the CP--properties of the Higgs bosons in Figs. 1-3, and the effective scalar-pseudoscalar-$Z$ boson vertex in (10) then
one can analyze the effective vertex $|\hat{V}_{Z H_{i} H_{j}}|$ by identifying the loop and tree-level contributions for each
$H_{i}$. For the parameter set in (12), the light stop mass start with $274\; \mbox{GeV}$ ($234\; \mbox{GeV}$) at $\gamma_{A}=0$,
and falls to  $177\; \mbox{GeV}$ ($227\; \mbox{GeV}$) at $\gamma_{A}=\pi$ for $\tan\beta=2$ (30). For $\tan\beta=2$,
$m_{\tilde{t}_{1}}$ falls below $250\; \mbox{GeV}$ at $\gamma_{A}\sim 1$. 

Depicted in Fig. 4 is the $\gamma_{A}$ dependence of $|\hat{V}_{Z H_{1} H_{3}}|$ for $\tan\beta=2$ (left panel) and $\tan\beta=30$
(right panel). Here solid curve shows $|V_{Z H_{1} H_{3}}|$ which includes no vertex corrections. For $\tan\beta=2$ (left panel) $|V_{Z
H_{1} H_{3}}|\sim 0.36$ at $\gamma_{A}=0$, and it gradually decreases with increasing  $\gamma_{A}$ eventually vanishing at
$\gamma_{A}=\pi$. This behaviour is dictated by the left panels of Figs. 1 and 3 where $H_{1}$ remains essentially CP--even for all
$\gamma_{A}$ whereas $H_{3}$ becomes a pure CP--even scalar in a narrow region around $\gamma_{A}=\pi$. The dashed curve shows 
$|\hat{V}_{Z H_{1} H_{3}}|$ when only the top quark loop is considered. The top quark loop is seen to contribute by $\sim 0.5 \%$
and has essentially the same $\gamma_{A}$ dependence as the solid curve. This essentially follows from the dominance of the 
`tree--vertex' $V_{Z H_{1} H_{3}}$. Here one notes that the absorbtive part of the top quark loop does not interfere with $V_{Z H_{1}
H_{3}}$ and becomes a two--loop effect in computing $|\hat{V}_{Z H_{1} H_{3}}|$. When the scalar top quark loops (dotted curve) are
included, however, one gets a $\sim 6\%$ enhancement in $|V_{Z H_{1} H_{3}}|$ at $\gamma=0$. This amount of enhancement is typical of
this process  as already noted in \cite{pokorski}. This large correction is due to the fact that $H_{i}^{S,P}$ to
${\tilde{t}_{1}}{\tilde{t}_{2}}$ couplings are large because of relatively large value of $|A_{t}|$. In fact, in the CP--conserving
limit $H_{i}^{P}$ couples only distinct stops so that the stop loop involves at leat one $\tilde{t}_{2}$. $|\hat{V}_{Z H_{1} H_{3}}|$
starts with this relatively large value at $\gamma_{A}=0$, and decreases faster than the previous two  
until $\gamma_{A}\sim 1.1$. This decrease follows from the gradual increase in $m_{\tilde{t}_{2}}$ suppressing the vertex formfactors. 
In this region $\gamma_{A}$ dependence are dictated by $R_{i j}$ and the vertices listed in (7)-(9). At $\gamma_{A}\sim 1.1$, however,
it changes sharply due to the fact that ${\cal{Q}}_{\tilde{t}}(-p_{i},p_{j},m_{\tilde{t}_{1}},m_{\tilde{t}_{1}},m_{\tilde{t}_{1}})$
now develops an absorbtive part (since at $\gamma_{A}\sim 1.1$ $m_{\tilde{t}_{1}}$ just falls below $\sqrt{s}/2$) which, when
multiplied by the purely imaginary  vertex $V_{H_{3}\tilde{t}_{1}\tilde{t}_{1}}^{(P)}$, becomes pure real and interferes with the
`tree-level' vertex $V_{Z H_{1} H_{3}}$ modifying it sharply. It is this kind of behaviour that shows clearly the effects of the 
pure light stop loop existing solely due to the CP--violation. Asymmetric behaviour of $|\hat{V}_{Z H_{1}H_{3}}|$ with respect to
$\gamma_{A}=\pi$ axis as well as its concidences with $|V_{Z H_{1} H_{3}}|$ follows from the $\gamma_{A}$ dependence of vertices
listed in (7)-(9): It behaves as $\sim \sin 2\gamma \cos 2 \gamma$ appropriate for $\gamma_{\mu}=0$ and $|A_{t}|> > |\mu|$.

The right panel of Fig. 4 shows the $\gamma_{A}$ dependence of $|\hat{V}_{Z H_{1} H_{3}}|$ for $\tan\beta=30$. One observes first
decrease in the strength of all curves compared to the left panel. One observes several differences between this figure and the 
left panel: First, as mentioned before, the light stop mass is below the $\sqrt{s}/2$ for the entire range of $\gamma$ so that 
there is no sharp change in the variation of the figure at any value of $\gamma_{A}$. Second, the comparatively fast variation of 
$|\hat{V}_{Z H_{1} H_{3}}|$ follows the right panels of Figs. 1 and 3 where $\phi_{2}$ component of $H_{1}$ remains large whereas 
the CP--odd component of $H_{3}$ is diminished rather fast (see the dotted curve in Fig. 3, right panel). It is for this reason that 
$|\hat{V}_{Z H_{1} H_{3}}|$ oscillates faster than in left panel. Third, $|\hat{V}_{Z H_{1} H_{3}}|$ for $\tan\beta$ is rescaled 
to smaller values compared to the left panel. This follows from the large value of $\tan\beta$ which is known to suppress tree
as well as the one--loop corrections \cite{pokorski}. However, the gain in $|\hat{V}_{Z H_{1} H_{3}}|$ with respect to its
value at $\gamma_{A}=0$ is much larger than the one in the left panel. For example, around $\gamma_{A}\sim 2.5$ $|\hat{V}_{Z H_{1}
H_{3}}|$ is $\sim 2.3$ times bigger than its value at $\gamma_{A}=0$. Finally one notices that, at large $\tan\beta$,
almost entire behaviour of $|\hat{V}_{Z H_{1} H_{3}}|$  is determined by the 'tree-vertex' $V_{Z H_{1} H_{3}}$; the one--loop 
corrections are suppressed compared to the left panel. In this limit top quark contribution is suppressed by $\cos\beta$ factor it
multiplies, on the other hand, stop contributions are suppressed by the $\cos\beta$ factor multiplying $|A|_{t}$ ($|\mu|$ is already
small) in $A_{P}^{i}$ and by similar factors in $A_{S}^{i}$. 
  
As is seen from Fig. 2 $H_{2}$ (which becomes $H$ at $\gamma_{A}=0$) is no longer a CP--even Higgs as $\gamma_{A}$ varies. Thus, it is
possible to produce $H_{1}H_{2}$ pairs in addition to $H_{1}H_{3}$ ditailed above. As expected, $|\hat{V}_{Z H_{1} H_{2}}|$ vanishes at
$\gamma_{A}=0$ at which both scalars are CP--even. In similarity with the discussions concerning $|\hat{V}_{Z H_{1} H_{3}}|$ above one
can analyze this process too. For example, minima and maxima of the effective vertex follow from their CP compositions in Figs. 2 and 3
using repestive panels. In general since the vertex is generated by the CP--violation in the Higgs sector, rather than the vertex
radiative corrections, behaviour of $|\hat{V}_{Z H_{1} H_{2}}|$ is mainly dictated by $V_{Z H_{1} H_{2}}$ everywhere. One finally
notices that the effects of the absorbtive part of the light stop contribution is too small to be seen in the variation of $V_{Z H_{1}
H_{2}}$.

In principle one can also analyze the effective vertex for $H_{2}H_{3}$ production. However, for the collider search the main
concern is the associated production of $H_{1}$ with a heavy Higgs scalar as this is the first step towards a complete discovery of
the Higgs spectrum of the MSSM. Moreover, for the center-of-mass energy chosen in this work pair production of such heavy scalars
will be suppressed. In spite of these facts, however, from the comparison of Figs. 2-3 it is obvious that $|\hat{V}_{Z H_{2}
H_{3}}|$ will not have a significant $\gamma_{A}$ dependence because these two scalars have complementary CP--properties. One
expects similar effects to occur in other Higgs boson couplings \cite{ben}.
\section{Discussions and Conclusion}
This work has concentrated on the associated production of one light and one heavy Higgs scalars in the presence of explicit
CP violation due to non-vanishing supersymmetric CP phases. In the light of the results derived in the text one can state that: 
$({\bf i})$ The lightest of the three Higgs scalars remains essentially CP--even  due to decoupling between the light and 
heavy sectors in the MSSM \cite{decoupling}, and this is also confirmed by \cite{ben,wagner}. $({\bf ii})$ Due to this CP--purity 
of the lightest Higgs $e^{+}e^{-}\rightarrow H_{i} H_{j}$ probe only the CP--odd composition of the heavy scalar; therefore, 
assuming sufficient mass resolution at next generation of colliders one expects MSSM to behave as having two CP--odd Higgs bosons 
for certain values of the supersymmetric CP--phases. $({\bf iii})$ Explicit CP--violation not only mixes scalars of different
CP--properties but also induces a novel vertex where a CP--odd Higgs boson can couple to identical sfermions. Effects of this
additional interaction rule has been shown to be observable by through the phase dependence of the 
one--loop scalar--pseudoscalar--$Z$ boson vertex. However, this additional interaction can show up most significantly in the
$pseudoscalar$ $\rightarrow \gamma \gamma$ decay whose rate gets no contribution from the sfermion loops in the absence of
CP--violation. $({\bf iv})$ The explicit CP--violation discussed here, in particular the relations among vertices coupling Higgs
scalars to gauge bosons and fermions, are special cases of the general rules described in \cite{gunion}. 
 
This work as well as others \cite{pilaftsis,ben,wagner} assume the existence of non-vanishing phases for $\mu$ parameter and all
other soft mass parameters. Though it is not essential for low-enegry supersymmetry phenomenology one can relate these phases to   
the Goldstone bosons of some broken global symmetries of the hidden sector in supergravity breaking \cite{sugra}. As the analysis
of \cite{phase2} shows these phases generally relax to (near) a CP--conserving point in the case of universality. However, in  
more general cases, especially when the supersymmetry breaking in the hidden sector is realized non-linearly, these phases may
relax to  some point away from the CP--conserving limit. In this sense, it is this possibility that is investigated here. As the
graphs of the effective vertices in Figs. 4-5 show, if these phases relax to points away from the CP--conserving point they can
have drastic implications for the collider phenomenology of Higgs bosons \cite{pilaftsis,ben,wagner}, and also for Kaon and  
B-meson phenomenology \cite{biz}. Finally one notes the acceleration in the interest to the supersymmetric CP--phases in the 
context of the electroweak baryogenesis \cite{baryo}.

\section{Acknowledgements}
The author is grateful to A. Masiero, A. Pilaftsis, J. Rosiek, G. Senjanovic and C. E. M. Wagner for stimulating discussions 
about various aspects of this work. He also thanks to CERN Theory Division where this work is completed.

\begin{figure}[htb]
\centerline{
\psfig{figure=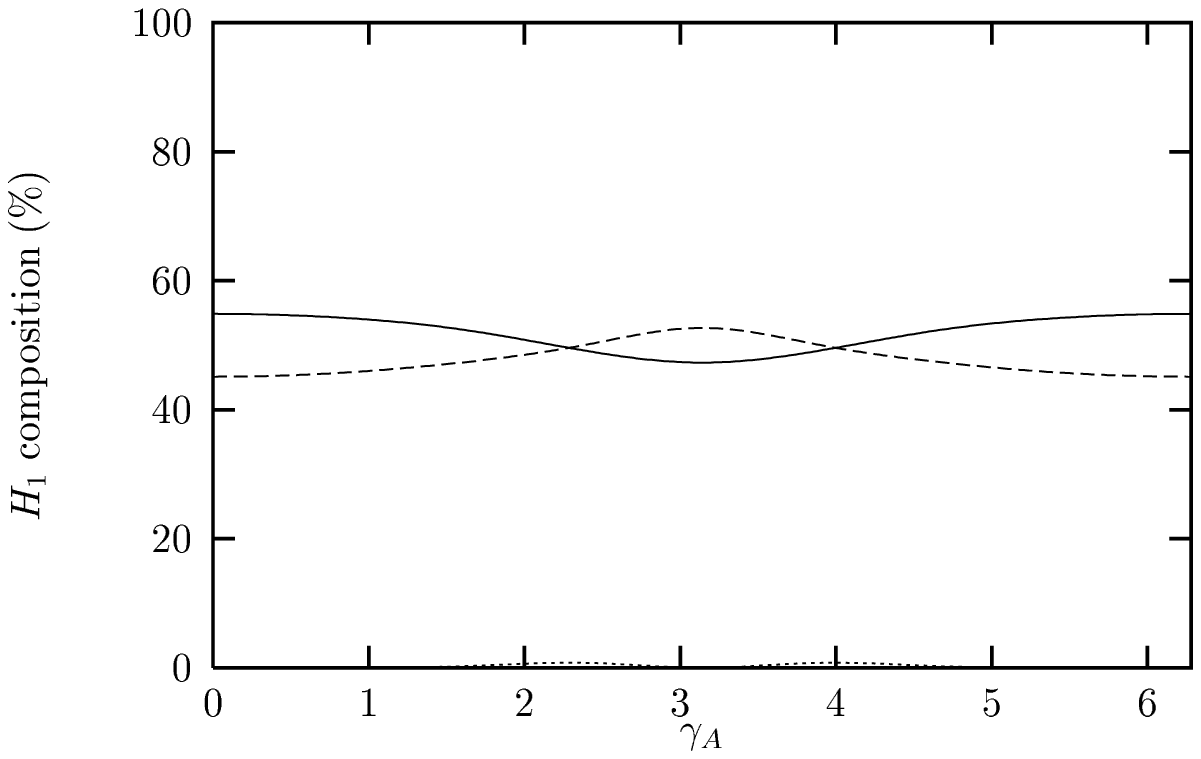,height=10cm,width=8cm,bbllx=-1.cm,bblly=6.cm,bburx=18.cm,bbury=21.cm}
\psfig{figure=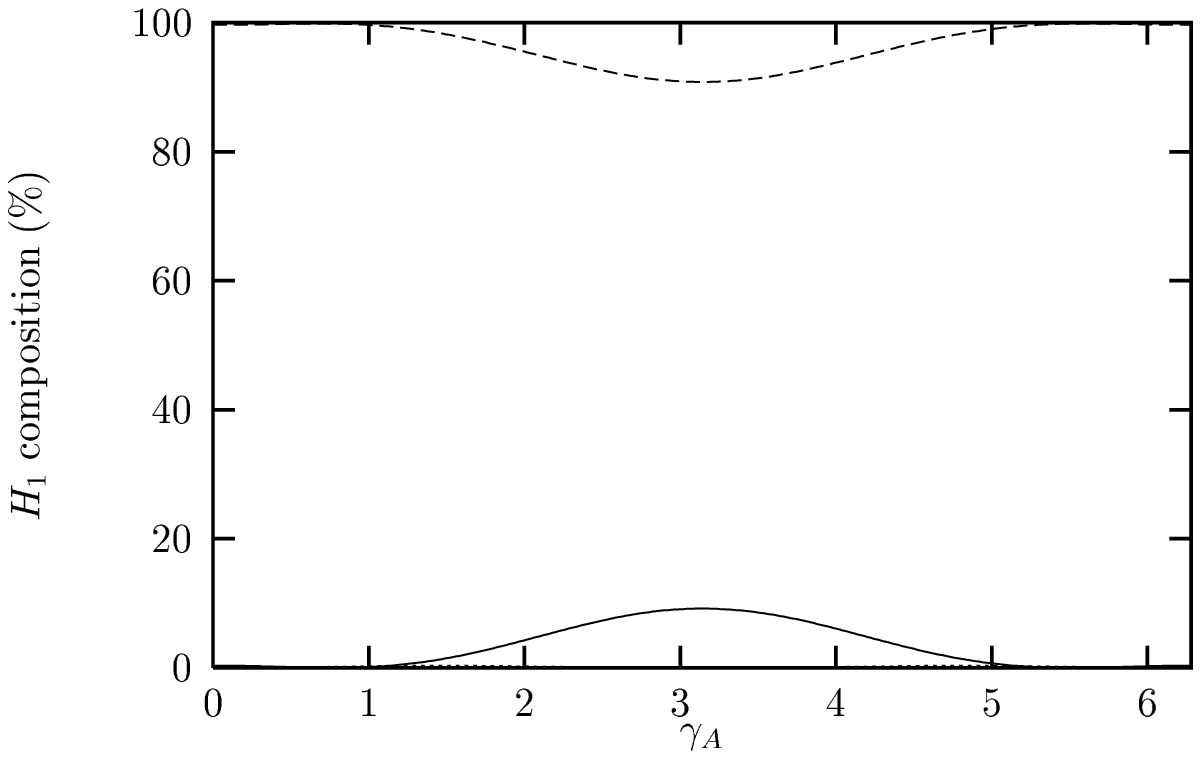,height=10cm,width=8cm,bbllx=1.cm,bblly=6.cm,bburx=20.cm,bbury=21.cm}}
\caption{\footnotesize Percentage composition of $H_{1}$ as a function of $\gamma_{A}$ for $\tan\beta=2$
(left panel) and $\tan\beta=30$ (right panel). Here $R_{11}^{2}$, $R_{12}^{2}$ and $R_{13}^{2}$ are 
shown by solid, dashed, and dotted curves, respectively.}
\end{figure}
\begin{figure}[htb]
\centerline{  
\psfig{figure=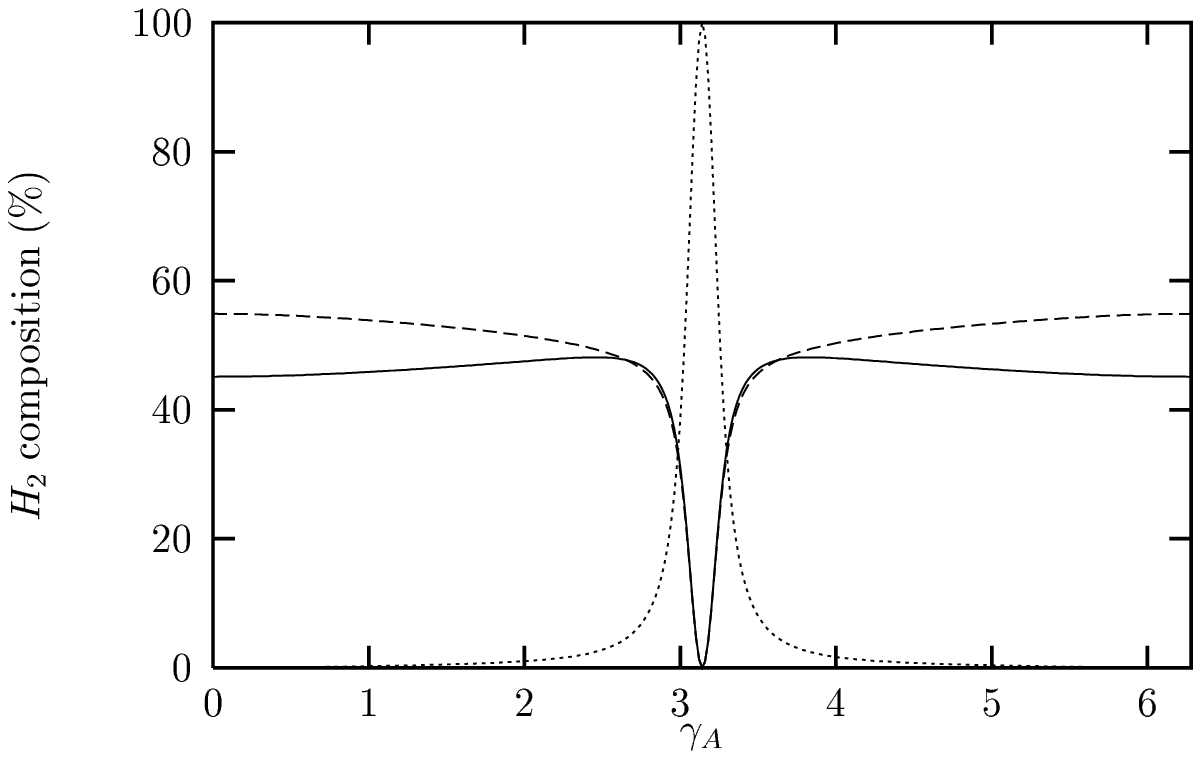,height=10cm,width=8cm,bbllx=-1.cm,bblly=6.cm,bburx=18.cm,bbury=21.cm}
\psfig{figure=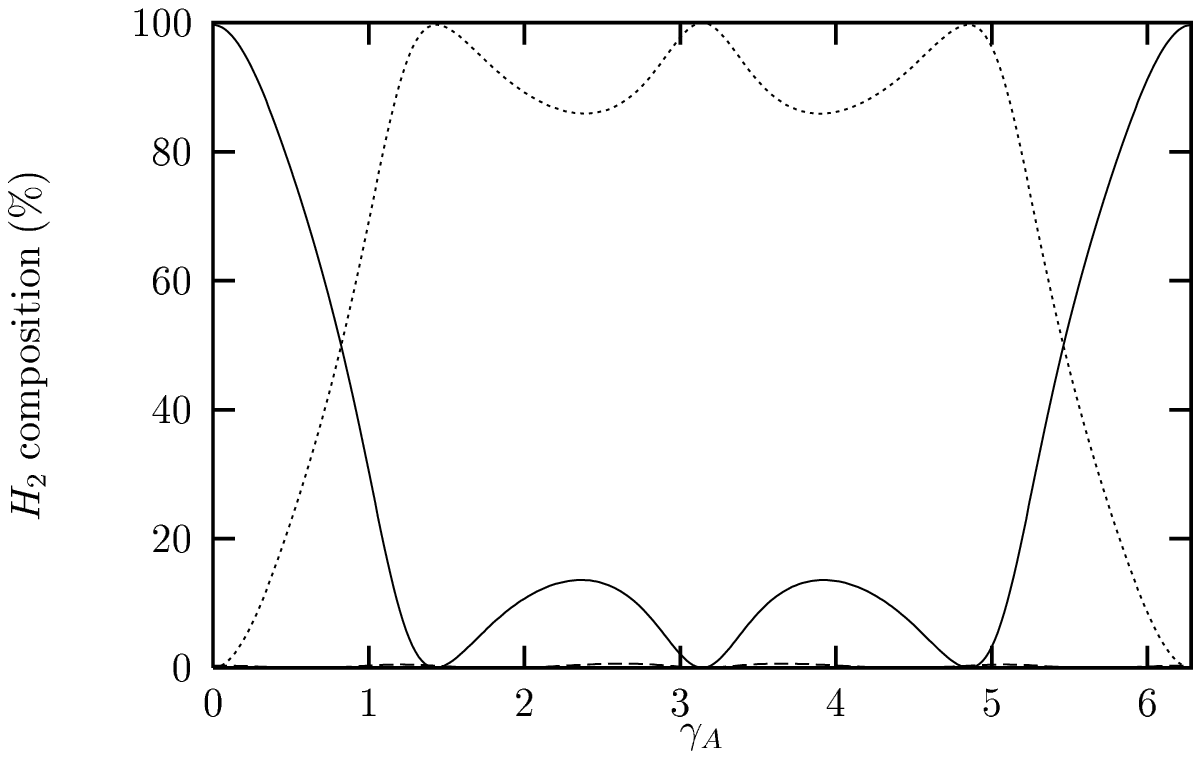,height=10cm,width=8cm,bbllx=1.cm,bblly=6.cm,bburx=20.cm,bbury=21.cm}}
\caption{\footnotesize Percentage composition of $H_{2}$ as a function of $\gamma_{A}$ for $\tan\beta=2$
(left panel) and $\tan\beta=30$ (right panel). Here $R_{21}^{2}$, $R_{22}^{2}$ and $R_{23}^{2}$ are
shown by solid, dashed, and dotted curves, respectively.}
\end{figure}
\begin{figure}[htb]
\centerline{  
\psfig{figure=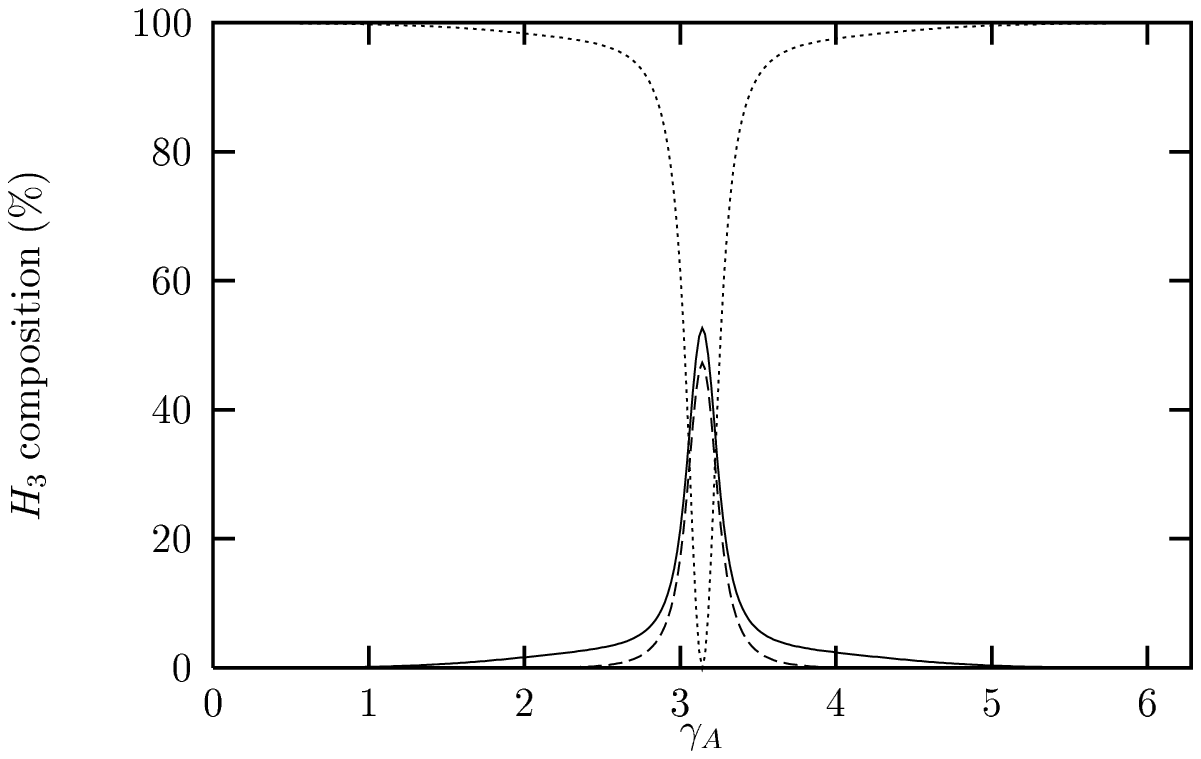,height=10cm,width=8cm,bbllx=-1.cm,bblly=6.cm,bburx=18.cm,bbury=21.cm}
\psfig{figure=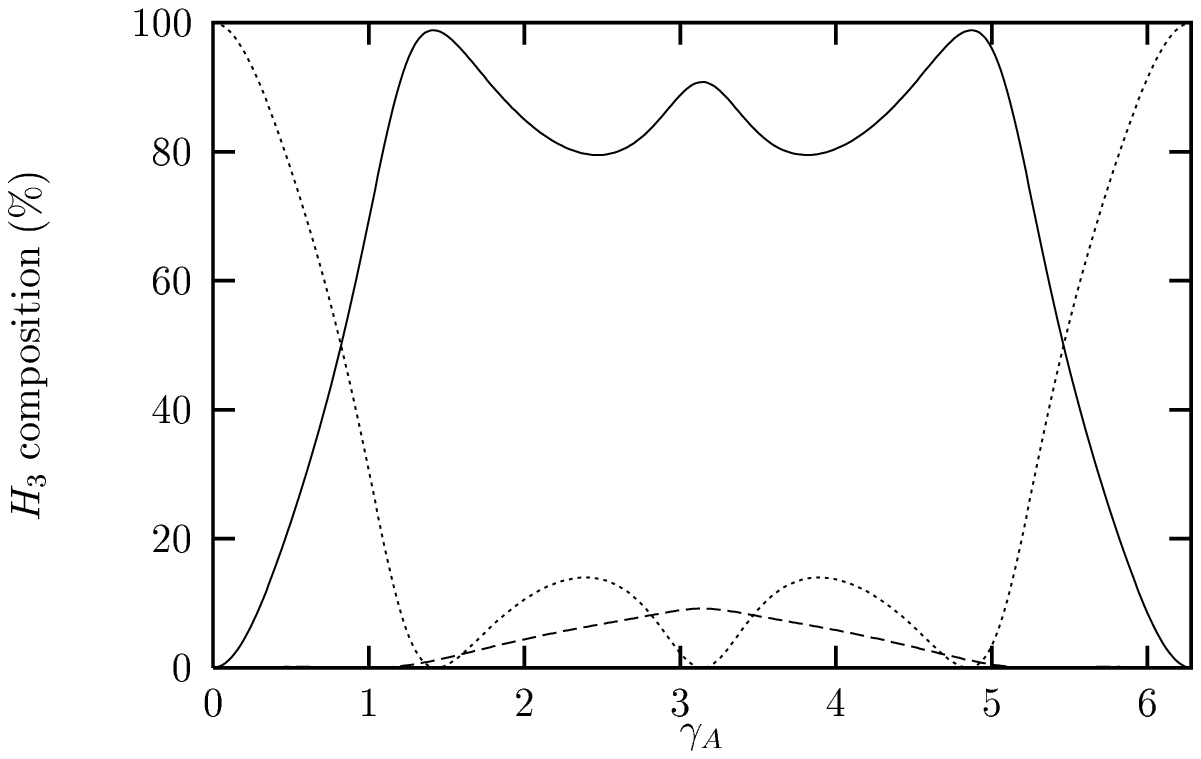,height=10cm,width=8cm,bbllx=1.cm,bblly=6.cm,bburx=20.cm,bbury=21.cm}}
\caption{\footnotesize Percentage composition of $H_{3}$ as a function of $\gamma_{A}$ for $\tan\beta=2$
(left panel) and $\tan\beta=30$ (right panel). Here $R_{31}^{2}$, $R_{32}^{2}$ and $R_{33}^{2}$ are
shown by solid, dashed, and dotted curves, respectively.}
\end{figure}
\begin{figure}[htb]
\centerline{
\psfig{figure=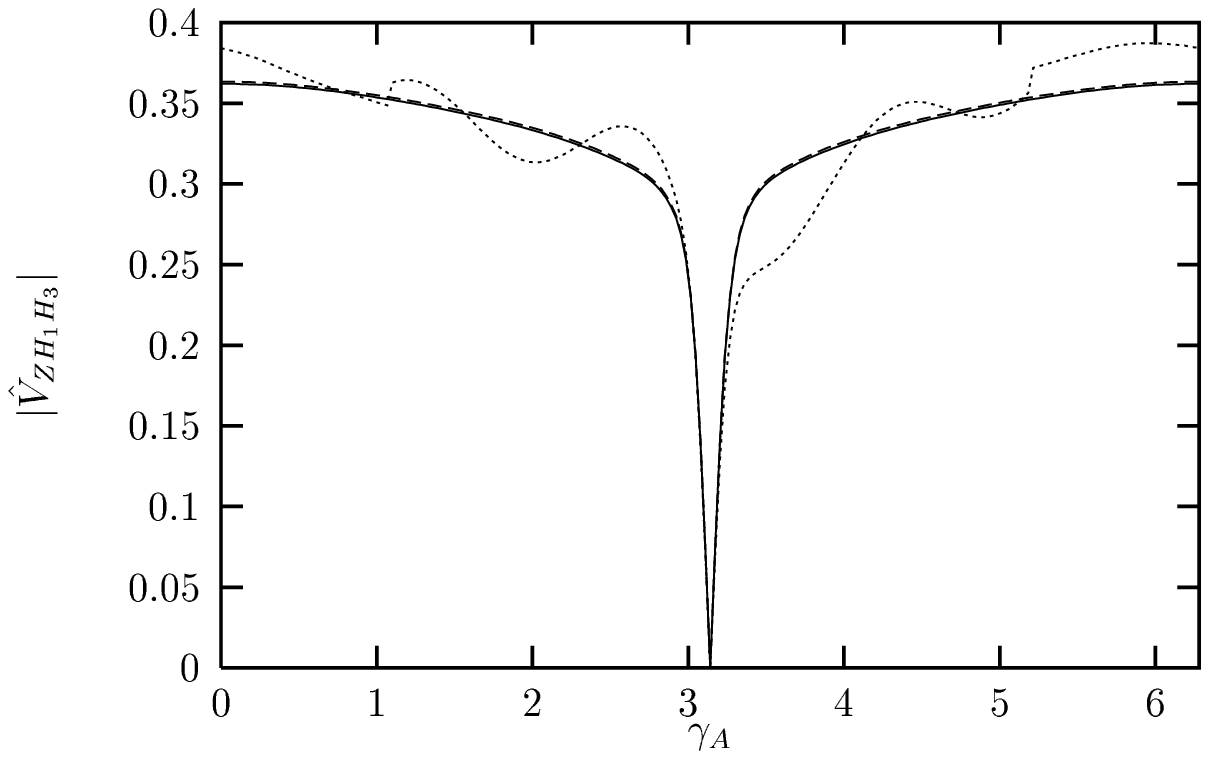,height=10cm,width=8cm,bbllx=-1.cm,bblly=6.cm,bburx=18.cm,bbury=21.cm}
\psfig{figure=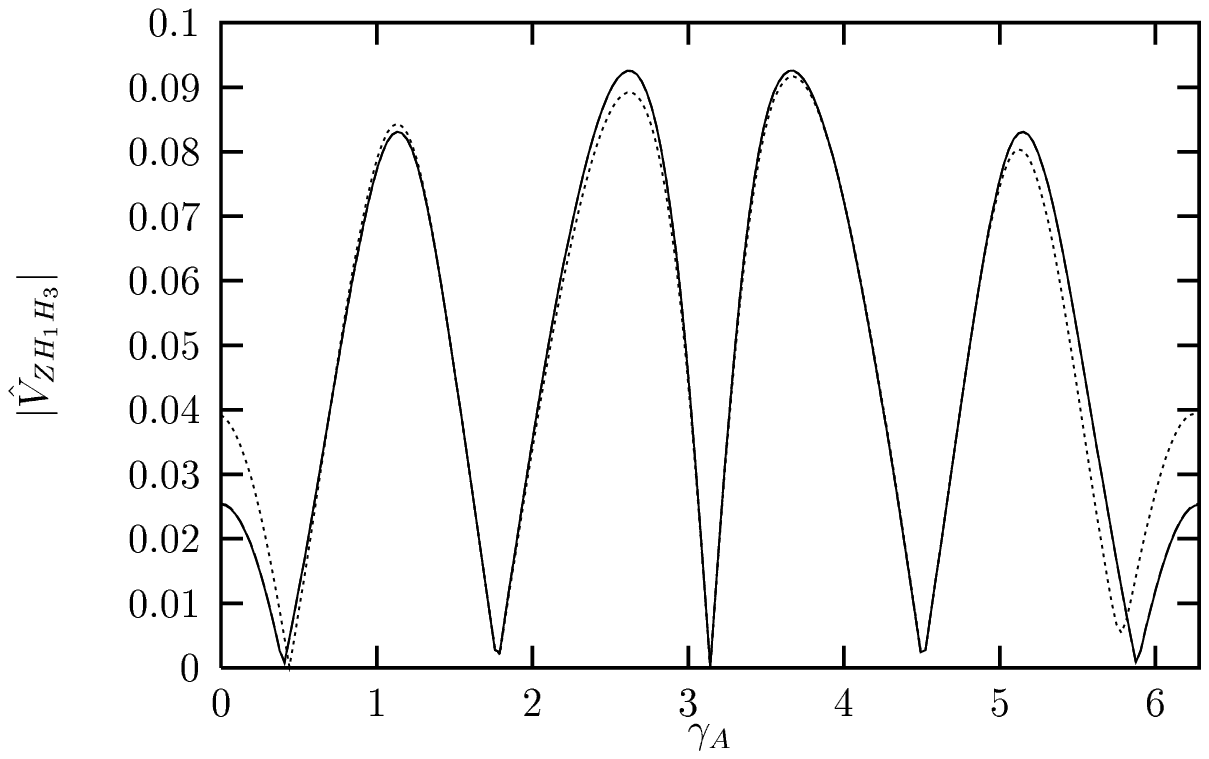,height=10cm,width=8cm,bbllx=1.cm,bblly=6.cm,bburx=20.cm,bbury=21.cm}}
\caption{\footnotesize Variation of $\hat{V}_{Z H_{1} H_{3}}$ with $\gamma_{A}$ when there is no vertex
corrections (solid curve), when only top quark contribution is included (dashed curve), and when both 
top quark and top squark loops are included (dotted curve). Here left panel stands for $\tan\beta=2$ and 
right panel for $\tan\beta=30$.}
\end{figure}
\begin{figure}[htb]
\centerline{
\psfig{figure=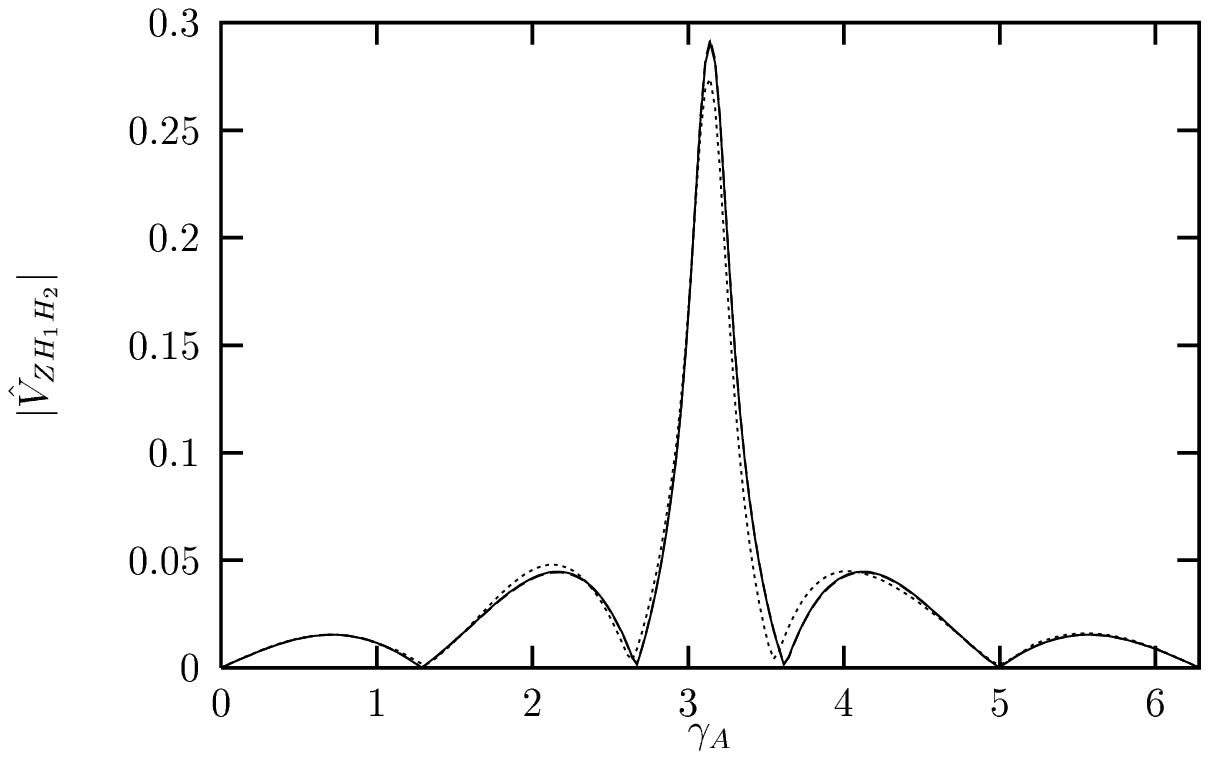,height=10cm,width=8cm,bbllx=-1.cm,bblly=6.cm,bburx=18.cm,bbury=21.cm}
\psfig{figure=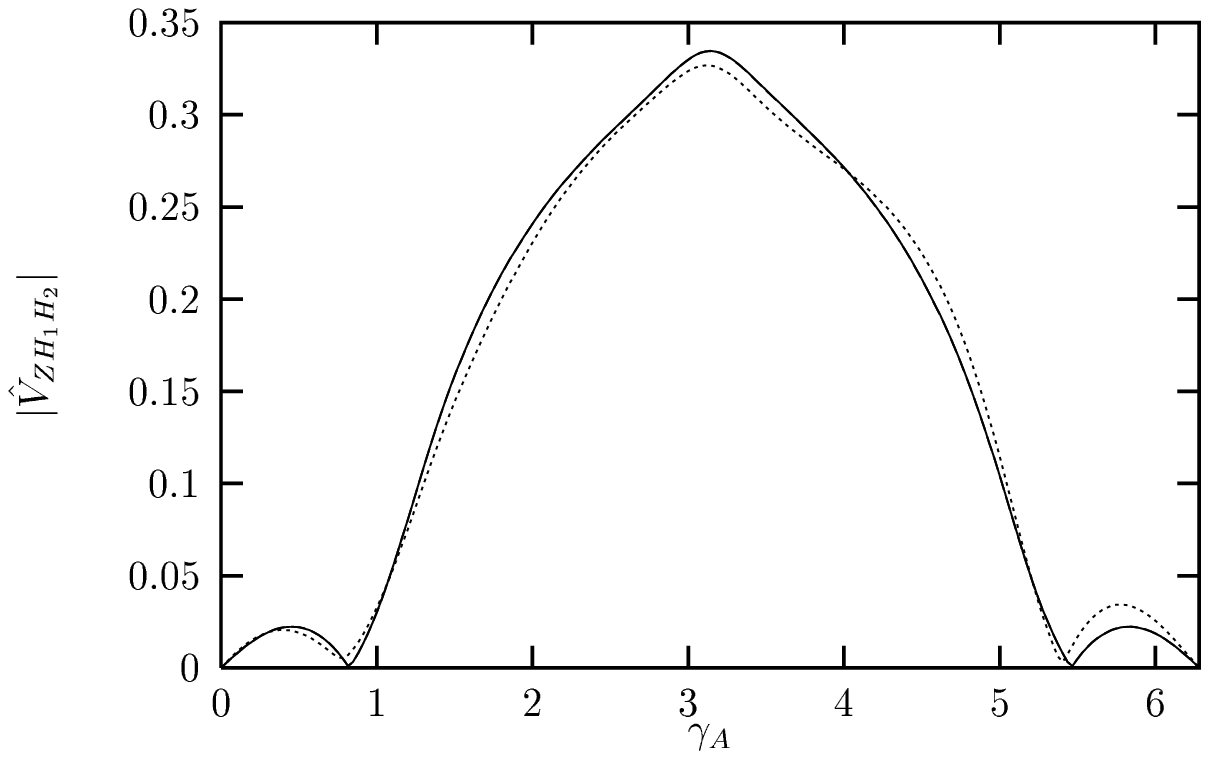,height=10cm,width=8cm,bbllx=1.cm,bblly=6.cm,bburx=20.cm,bbury=21.cm}}
\caption{\footnotesize Variation of $\hat{V}_{Z H_{1} H_{2}}$ with $\gamma_{A}$ when there is no vertex
corrections (solid curve), when only top quark contribution is included (dashed curve), and when both
top quark and top squark loops are included (dotted curve). Here left panel stands for $\tan\beta=2$ 
and right panel for $\tan\beta=30$.}
\end{figure}
\end{document}